\documentclass[12pt]{article}

\usepackage{amsfonts,amssymb}
\usepackage{tikz}

\renewcommand{\ker}{{\rm ker}}
\newcommand{\im }{{\rm im}}

\newcommand{\Hom}{{\rm Hom}}
\newcommand{\ra}{{\rightarrow}}

\newcommand{\ZZ}{{\mathbb Z}}
\newcommand{\RR}{{\mathbb R}}

\newcommand{\frg}{{\mathfrak g}}
\newcommand{\frh}{{\mathfrak h}}
\newcommand{\dt}{{\bar t}}

\newcommand{\U}{{\mathfrak U}}
\newcommand{\cc}{{\cup_1}}
\newcommand{\frp}{{\mathfrak P}}

\newcommand{\Tr}{{\rm Tr}}

\author{Anton Kapustin \\ {\it California Institute of Technology, Pasadena, CA} \and Ryan Thorngren \\ {\it Department of Mathematics, University of California, Berkeley, CA}}

\title{Topological Field Theory on a Lattice, Discrete Theta-Angles and Confinement}

\begin{document}

\titlepage
\maketitle

\abstract{We study a topological field theory describing confining phases of gauge theories in four dimensions. It can be formulated  on a lattice using a discrete 2-form field talking values in a finite abelian group (the magnetic gauge group). We show that possible theta-angles in such a theory are quantized and labeled by quadratic functions on the magnetic gauge group. When the theta-angles vanish, the theory is dual to an ordinary topological gauge theory, but in general it is not isomorphic to it. We also explain how to couple a lattice Yang-Mills theory to a TQFT of this kind so that the 't Hooft flux is well-defined, and quantized values of the theta-angles are allowed. The quantized theta-angles include the discrete theta-angles recently identified by Aharony, Seiberg and Tachikawa.}

\section{Introduction}

In this paper we use TQFT to study massive phases of gauge theories in four dimensions. It is believed that pure Yang-Mills theory with a compact semi-simple gauge group $G$ is confining and has a mass gap. However, this does not mean that the long-wavelength behavior of the theory is necessarily trivial. Rather, it is expected to be described by a unitary 4d TQFT, and one would like to identify this TQFT.  In this paper we study such TQFTs in detail, paying special attention to topological terms. More generally, one might consider massive phases of gauge theories where the gauge group is partially spontaneously broken and partially confined. Such theories are more complicated and will be discussed elsewhere \cite{two}.

If the microscopic gauge group is $G$, one of the fields of the TQFT should be a gauge field which is locally a 1-form with values in the Lie algebra $\frg$ of $G$.
It was argued in \cite{GK} that a TQFT which describes the confining phase should also involve a nonabelian B-field which is locally a 2-form with values in $\frg$.
The TQFT is not determined by $G$; one also needs to choose a cover $H$ of $G$ and specify theta-angles. The theta-angles are discrete, i.e. satisfy a quantization condition.\footnote{If $G$ is simply-connected, the TQFT describing the confining phase is trivial.}

We show first of all that the confining TQFT depends only on the kernel of the covering homomorphism $t: H\ra G$ and the theta-angles. We will refer to $\Pi_2=\ker\  t$ as the magnetic gauge group. The allowed values of the theta-angles are determined by $\Pi_2$ alone. Such a theory was previously discussed in the mathematical literature \cite{Quinn, FHLT}, but the connection with ordinary gauge theories was not noted.

We provide a rigorous lattice formulation of the confining TQFT. The only fields in the lattice formulation is a discrete 2-form taking values in $\Pi_2$. 
If the theta-angles vanish, one can dualize it to a discrete gauge field with gauge group $\Pi_2^*=\Hom(\Pi_2,U(1))$, the Pontryagin dual of $\Pi_2$. This is a manifestation of electric-magnetic duality. However, in general the B-field cannot be dualized to an ordinary gauge field, and the confining TQFT is not equivalent to a topological gauge theory. Indeed, its partition sum depends on the signature of the underlying 4-manifold, while topological gauge theory, by definition, depends only on its fundamental group. 

Finally we address the microscopic origin of various confining phases. We write down a lattice formulation of Yang-Mills theory with quantized theta-angles which precisely correspond to the theta-angles in the low-energy TQFT. We also make contact with the work of Aharony, Seiberg and Tachikawa \cite{AST} who classified theta-angles, both discrete and continuous, in the continuum Yang-Mills theory. In particular, these authors showed that discrete theta-angles label the ambiguity in the choice of the spectrum of the allowed line operators. We find that all discrete theta-angles identified in \cite{AST} have a counterpart in lattice Yang-Mills theory. On the other hand, there is no good lattice counterpart of the instanton number, and accordingly no satisfactory lattice counterpart of the continuous theta-angle.

A.K. would like to thank Dan Freed, Sergei Gukov, Michael Hopkins, Nathan Seiberg, Yuji Tachikawa, and Constantin Teleman for discussions. R.T. would also like to thank Scott Carnahan, Evan Jenkins, Alex Rasmussen, David Roberts, and Urs Schreiber for discussions.
This work was supported in part by the DOE grant DE-FG02-92ER40701 and by the National Science Foundation under Grant No. PHYS-1066293 and the hospitality of the
Aspen Center for Physics.

\section{Confining TQFT in the continuum}

Let us recall the formulation of the TQFT describing a confining phase of gauge theory with a microscopic gauge group $G$ \cite{GK}. Let $H$ be a finite cover of $G$, $t:H\ra G$ be a covering map. The group $\ker\ t=\Pi_2$ is a subgroup of the center of $H$. Thus we have a well-defined action of $G$ on $H$ via
$$
g: h\mapsto \alpha(g)(h)=\tilde g h \tilde g^{-1}.
$$
Here $\tilde g$ is an element of $H$ satisfying $t(\tilde g)=g$. Although $\tilde g$ is not defined uniquely, the element $\alpha(g)(h)$ is well-defined and depends smoothly on $g$ and $h$. Note that $t$ identifies the Lie algebras of $H$ and $G$, but we will not identify them in our notation and will denote them $\frh$ and $\frg$ respectively. We will denote by $\dt$ the isomorphism $\frh\ra\frg$. 

The triple $(G,H,t)$ encodes the topological charges of the monopole condensate \cite{GK}. Namely, while the microscopic gauge theory has 't Hooft flux taking values in $\pi_1(G)$, the TQFT is designed so that conserved 't Hooft flux takes values in the quotient $\pi_1(G)/\pi_1(H)=\Pi_2.$ The interpretation is that the 't Hooft fluxes of monopoles in the condensate generate the subgroup $\pi_1(H)$, and accordingly the 't Hooft flux at long distances is conserved only modulo elements of $\pi_1(H)$. For example, if $G=SU(N)/\ZZ_N$, two natural choices for $H$ are $H=SU(N)$ and $H=SU(N)/\ZZ_N$. In the first case, $\pi_1(H)=0$, so the monopole condensate consists only of monopoles with a trivial 't Hooft flux (but nontrivial GNO flux \cite{GNO,KapWH}). In the second case, monopoles with the miminal 't Hooft flux are present in the condensate, and at long distances no conserved 't Hooft flux can be defined at all, i.e. $\Pi_2=0$.

The fields of the TQFT are locally a $\frg$-valued 1-form $A$ and an $\frh$-valued 2-form $B$. If the space-time manifold $X$ is $\RR^n$ and the fields are everywhere non-singular, this also applies globally, but in general, to specify a field configuration,  one needs to choose a good open cover $\U=\left\{ U_i, i\in I\right\}$ of $X$ and specify both the fields on each $U_i$  and transitions functions on double and triple overlaps. Namely, on each $U_i$ we have an $\frg$-valued 1-form $A_i$ and an $\frh$-valued 2-form $B_i$, on each $U_{ij}=U_i\bigcap U_i$ we have a $G$-valued function $g_{ij}$ and an $\frh$-valued 1-form $\lambda_{ij}$, and on each $U_{ijk}=U_i\bigcap U_j\bigcap U_k$ we have an $H$-valued function $h_{ijk}$. These data should satisfy the following compatibility conditions. On each $U_{ij}$ we must have
\begin{eqnarray}
A_j & = & g_{ij} A_i g_{ij}^{-1}+g_{ij} dg_{ij}^{-1}-\dt(\lambda_{ij}),\\
B_j & = & {\tilde g}_{ij} B_i  {\tilde g}_{ij}^{-1}-d_{A_j}\lambda_{ij}-\lambda_{ij}\wedge \lambda_{ij}.
\end{eqnarray}
On each $U_{ijk}$ we must have 
\begin{eqnarray}
h_{ijk}^{-1}\lambda_{ik} h_{ijk} & = & {\tilde g}_{jk}\lambda_{ij}{\tilde g}_{jk}^{-1}+\lambda_{jk} -h_{ijk}^{-1} d h_{ijk} %\nonumber \\
 -h_{ijk}^{-1}\, \dt^{-1}(A_k) h_{ijk} + \dt^{-1} (A_k) \,. \nonumber
\end{eqnarray}
and
$$
g_{ik}=t(h_{ijk})g_{jk} g_{ij}.
$$
On each $U_{ijkl}=U_i\bigcap U_j\bigcap U_k\bigcap U_l$ we must have
$$
h_{ijl}h_{jkl} \, = h_{ikl} \cdot {\tilde g}_{kl} h_{ijk}{\tilde g}_{kl}^{-1}.
$$
The action is
$$
\int (F_A-\dt(B))\wedge b+\ldots
$$
where $b$ is locally a 2-form with values in $\frg$,  $F_A=dA+A\wedge A$, and dots denote topological terms multiplied by theta-angles. In \cite{GK} these terms were written as 
$$
\int \langle F_A,\wedge F_A\rangle,
$$
but this is not a well-defined expression since $F_A$ does not transform homogeneously as one goes from $U_i$ to $U_j$ and therefore is not a 2-form with values in a vector bundle. We will specify the topological terms more precisely below. On the other hand, the combination $F_A-\dt(B)$ transforms homogeneously:
$$
F_{A_j}-\dt(B_j)=g_{ij} (F_{A_i}-\dt(B_i)) g_{ij}^{-1},
$$
 so the equation of motion $F_A-\dt(B)=0$ is gauge-invariant. The equation of motion allows one to solve for $B$ in terms of $A$, which is what we will do. 

Gauge transformations are parameterized by a collection of $G$-valued functions $g_i$ and $\frh$-valued 1-forms $\lambda_i$ on each $U_i$ and a collection of $H$-valued functions $h_{ij}$ on each $U_{ij}$. Their detailed form is described in \cite{GK}. We can gauge away $A_i$ by a gauge transformation with $g_i=1, \lambda_i=A_i$ and $h_{ij}=1$. Then we can gauge away $\lambda_{ij}$ by a gauge transformation $g_i=1, \lambda_i=0,$ and $h_{ij}=g_{ij}^{-1}$ (this second step is only possible because $t$ is surjective). After this the only datum left is a collection of functions $h_{ijk}: U_{ijk}\ra \Pi_2$ satisfying a cocycle condition on each $U_{ijkl}$:
$$
h_{ijl}h_{jkl} \, = h_{ikl} h_{ijk}.
$$
Following a long-standing tradition, we will denote groups of $p$-cochains by $C^p$ and groups of $p$-cocycles by $Z^p$. Thus $h$ is an element of the group of Cech 2-cocycles $Z^2(\U,\Pi_2)$.
 Residual gauge transformations are parameterized by functions $h_{ij}: U_{ij}\ra \Pi_2$, i.e. by elements of the group $C^1(\U,\Pi_2)$. They change the cocycle $h$ by a coboundary. There are also gauge transformations between gauge transformations; they are parameterized by functions $h_i:U_i\ra \Pi_2$, i.e. by elements of $C^0(\U,\Pi_2)$. We will refer to them as 2-gauge transformations.

The partition sum thus reduces to a sum over elements of $Z^2(\U;\Pi_2)$. The normalization factor is one over the order of the group of gauge transformations $C^1(\U,\Pi_2)$ times the order of the group of 2-gauge transformations $C^0(\U,\Pi_2)$.

The topological term in the action must be an integral of an element of $H^4(\U,\RR/\ZZ)$ over the fundamental 4-cycle $[X]$.  Its construction is discussed in the next section.

Observables in this TQFT are very simple. Given a character of $\Pi_2$ and a 2-cycle $\Sigma$ in $X$, we can construct a Wilson surface observable by evaluating the 2-cocycle $h$ on $\Sigma$ and then evaluating the character on the resulting element of $\Pi_2$. The Wilson surface observable is obviously gauge-invariant.  In \cite{GK} such observables were referred to as electric surface operators. There are also defect loops (essentially 't Hooft loops) labeled by an element $w\in\Pi_2$ and a homologically trivial 1-cycle $\gamma$. Such a defect is defined by the condition that on any $S^2$ which has a linking number $1$ with $\gamma$ the class $h$ evaluates to $w$. 

\section{Lattice formulation of the TQFT}

In this section and henceforward we write $\Pi_2$ additively and let $U(1)=\mathbb{R}/\mathbb{Z}$.

The lattice formulation of the TQFT is mostly concretely given in terms of a triangulation $K$ of the space-time $X$. The triangulation needs to be made of oriented simplices. That is, the 0-simplices are ordered. However, it is also possible to give a manifestly topological description of this theory similar to that of the Dijkgraaf-Witten gauge theory \cite{DW}, so it does not actually depend on the triangulation. First, we give the concrete description.

A configuration is an assignment of elements $h_\Sigma \in \Pi_2$ to each 2-simplex $\Sigma$, subject to the constraint that this assignment is flat. This means that for every 3-simplex with its 2-simplices assigned $h_0,h_1,h_2,h_3$ we require
$$
h_0-h_1+h_2-h_3 = 0,
$$
where these elements are labeled according to the ordering on the 3-simplex and the standard notation convention for face maps (that is, $h_i$ denotes the value of the 2-cocycle on the face of the 3-simplex obtained by dropping the $i^{\rm th}$ vertex). Equivalently, the sum is made with signs corresponding to the orientation of each face relative to the 3-simplex.

 A gauge transformation is an assignment of elements $f_\gamma \in \Pi_2$ to each 1-simplex $\gamma$. If the boundary of a 2-simplex $\Sigma$ is assigned $f_0,f_1,f_2$, then
 $$
 h_\Sigma \mapsto h_\Sigma + f_0 - f_1 + f_2,
 $$
 with the sign conventions as above.
 
 There are also 2-gauge transformations. These are parametrized by an assignment of $m_p\in \Pi_2$ to every 0-simplex $p$. If $\partial\gamma$ is assigned $m_0, m_1$, then $f_\gamma$ transforms as follows:
 $$
 f_\gamma \mapsto f_\gamma + m_0 - m_1.
 $$
 One easily checks that gauge transformation related by a 2-gauge transformation act identically on configurations.
 
 The configuration data are equivalent to a simplicial cocycle $h \in Z^2(K,\Pi_2)$. Gauge transformations shift $h$  by the differential of a 1-cochain, and 2-gauge transformations shift those 1-cochains by the differential of a 0-cochain. Gauge equivalence classes are thus cohomology classes in $H^2(K,\Pi_2)$. It is well known that such classes are equivalent to homotopy classes of maps $h:X\to B^2\Pi_2$, where we write $B^2\Pi_2$ for the Eilenberg-MacLane space $K(\Pi_2,2)$ with the 2nd homotopy group $\Pi_2$ and the rest vanishing \cite{Hatcher}. 
 
 The action functional is defined by a class $\mathcal{L} \in H^4(B^2\Pi_2,\RR/\ZZ)$ as follows:
 $$
 S(h)= 2\pi i \ \int_X h^*\mathcal{L},
 $$
 where $h:X \to B^2\Pi_2$ is the classifying map of the configuration. Gauge transformations are homotopies of this map, as explained above, so the resulting theory is topological (and even homotopy-invariant). 
 
It is possible give a very concrete description of this action. By the universal coefficient theorem and the vanishing of $H_3(B^2\Pi_2,\ZZ)$ \cite{EM}, we get
$$
H^4(B^2\Pi_2,U(1))=\Hom(H_4(B^2\Pi_2,\ZZ),U(1)).
$$
Now, according to \cite{EM}, for any abelian group $\Pi$ one has $H_4(B^2\Pi,\ZZ)=\Gamma(\Pi)$, where $\Gamma(\Pi)$ is the universal quadratic group for $\Pi$.  To explain what this means, recall that a quadratic function on an abelian group $\Pi$ with values in an abelian group $A$ is a map $q:\Pi\ra A$ satisfying two properties: $q(-x)=q(x)$ for any $x\in \Pi$, and $b(x,y)=q(x+y)-q(x)-q(y)$ is a bilinear function from $\Pi\times\Pi$ to $A$. To classify quadratic functions on $\Pi$ with values in various $A$, it is convenient to introduce an abelian group $\Gamma(\Pi)$ equipped with a quadratic function $\gamma:\Pi\ra \Gamma(\Pi)$ with the following property. For any abelian group $A$ any quadratic function on $\Pi$ with values in $A$ can be presented as a composition of $\gamma$ and a homomorphism from $\Gamma(\Pi)$ to $A$. Such a group $\Gamma(\Pi)$ exists and is uniquely defined by this property \cite{Whitehead_certain}. We call it the universal quadratic group for $\Pi$.

The group $\Gamma(\Pi)$ is easily computable for any finite abelian group $\Pi$. Namely, one can show that for $\Pi=\ZZ_r$ with $r$ odd $\Gamma(\Pi)=\ZZ_r$, while for $r$ even one has $\Gamma(\ZZ_r)=\ZZ_{2r}$ \cite{Whitehead_certain}. Once this is known, we can compute $\Gamma(\Pi)$ for any other finite abelian $\Pi$ using the property:
\begin{equation}\label{cross}
\Gamma(\oplus_i A_i)=\bigoplus_i \Gamma(A_i)\oplus \bigoplus_{i<j} A_i\otimes A_j. 
\end{equation}
The tensor product term can be interpreted as the universal source for bilinear maps from $A_i \times A_j$. This explains the above isomorphism. For example $\Gamma(\ZZ_2\oplus \ZZ_2)=\ZZ_4\oplus\ZZ_4\oplus\ZZ_2$.

Returning to our problem, we see that possible actions are classified by elements of 
$$
H^4(B^2\Pi_2,U(1))=\Hom(\Gamma(\Pi_2),U(1)),
$$
and the latter group can be identified with the group of quadratic $U(1)$-valued functions on $\Pi_2$. This result classifies possible discrete theta-angles in our TQFT. For example, we immediately get that if $\Pi_2=\ZZ_r$ with odd $r$, then theta-angles take values in $\ZZ_r$, while if $\Pi_2=\ZZ_r$ with even $r$, theta-angles take values in $\ZZ_{2r}$. 

We still need a concrete recipe to produce an action from a $U(1)$-valued quadratic function on $\Pi_2$. Such a recipe can be obtained as follows. As explained above, for any abelian group $A$ we have an isomorphism
$$
H^4(B^2\Pi_2,A)=\Hom(\Gamma(\Pi_2),A).
$$
Let us set $A=\Gamma(\Pi_2)$. The group $\Hom(\Gamma(\Pi_2),\Gamma(\Pi_2))$ has a distinguished element, the identity. Thus there should be a distinguished element in $H^4(B^2\Pi_2,\Gamma(\Pi_2))$. This distinguished element is known as the Pontryagin square \cite{Whitehead_polyhedra,Whitehead_certain} and is denoted $\frp$. Its definition and properties are discussed in the Appendix. One can think of $\frp$ as a cohomology operation, a functorial way to associate to $h\in H^2(K,\Pi_2)$ an element $\frp h\in H^4(K,\Gamma(\Pi_2))$, for any simplicial complex $K$. By definition, $\frp h=h^*\frp$, where we think of $h$ as a map to the classifying space $B^2\Pi_2$. 

Suppose that we are given a quadratic function $q:\Pi_2\ra U(1)$. Then the value of the action on the configuration $h$ is an integral over $X$ of
$$
2\pi i \ h^*\mathcal{L}=2\pi i \ h^*(q_*(\frp))=2\pi i \  q_*(\frp h),
$$
where $q_*:\Gamma(\Pi_2) \to \RR/\ZZ$ is the group homomorphism induced by $q$. Thus we obtain a concrete formula for the action provided we have a concrete formula for the Pontryagin square. These formulas are given in the appendix.

\section{Partition sum and duality}\label{partition}

In the case when the discrete theta-angles vanish, $q=0$, it is easy to evaluate the partition sum of the confining TQFT for any\footnote{Every smooth manifold admits a triangulation. If $X$ is only a topological 4-manifold, then it may not admit a triangulation. In that case we simply use the Cech complex instead of the simplicial complex to evaluate the partition sum.} closed 4-manifold $X$ (it does not matter in this case whether $X$ is orientable or not). Let $K$ be a triangulation of $X$. Since each configuration $h\in Z^2(K,\Pi_2)$ has weight one, we get
$$
Z_X(\Pi_2,0)=\frac{|H^2(K,\Pi_2)||H^0(K,|\Pi_2)|}{|H^1(K,\Pi_2)|}=|\Pi_2|^{e_X} \frac{|H^1(X,\Pi_2)|}{|\Pi_2|},
$$
where $|A|$ denotes the order of a finite abelian group $A$, and $e_X$ is the Euler characteristic of $X$. 

The factor $|\Pi_2|^{e_X}$ can be removed by adding to the action of the TQFT a purely geometric term $e_X \log|\Pi_2|$. Such a term is local (because Euler characteristic can be written as an integral of the Euler density). The TQFT action is usually regarded as defined modulo such terms, because their addition does not affect anything but the value of the partition sum on a closed 4-manifold (i.e. the vector spaces attached to closed 3-manifolds, the categories attached to closed 2-manifolds, etc., are not affected).  On the other hand, the factor
$$
 \frac{|H^1(X,\Pi_2)|}{|\Pi_2|}
$$
is physically significant and coincides with the partition sum of a topological gauge theory (4d Dijkgraaf-Witten theory) with gauge group $\Pi_2$ and  a trivial topological action. 

In fact, it is easy to show that for $q=0$ the confining TQFT is equivalent to the topological gauge theory whose gauge group is the Pontryagin dual of $\Pi_2$, i.e. $\Pi_2^*=\Hom(\Pi_2,\RR/\ZZ).$ The equivalence arises from a topological version of electric--magnetic duality. We can rewrite the partition sum of the confining TQFT as a sum over all cochains in $C^2(K,\Pi_2)$, at the expense of introducing a Lagrange multiplier field $g$ with values in $C^1(K^*,\Pi_2^*)$,
where $K^*$ is the dual cell complex of the triangulation $K$ (its cells are barycentric stars of the complex $K$, see e.g. \cite{Hatcher}).  The non-degenerate intersection pairing between $C^3(K,\Pi_2)$ and $C^1(K^*,\Pi_2^*)$ allows us to write the partition sum as
$$
Z_X(\Pi_2,0)=\frac{|C^0(K,\Pi_2)|}{|C^1(K,\Pi_2)||C^1(K^*,\Pi_2^*)|} \sum_{h,g} \exp(2\pi i \langle g, \delta h\rangle).
$$
where the summation is over $C^2(K,\Pi_2)\times C^1(K^*,\Pi_2^*)$. Performing summation over $C^2(K,\Pi_2)$ we get 
$$
Z_X(\Pi_2,0)=|\Pi_2|^{e_X} \frac{1}{|C^1(K^*,\Pi_2^*)|}\sum_{g\in Z^1(K^*,\Pi_2^*)} 1.
$$
Up to a factor $|\Pi_2|^{e_X}$ this is the partition sum of the topological gauge theory with gauge group $\Pi_2^*$, computed using the dual cell complex $K^*$.

If the discrete theta-angles do not vanish, the duality cannot be performed, and the theory is not equivalent to a topological gauge theory. Consider the other extreme where the quadratic form $q:\Pi_2\ra \RR/\ZZ$ is non-degenerate, i.e. the corresponding bilinear form is an isomorphism of $\Pi_2$ and $\Pi_2^*$. Then the partition sum is  $|\Pi_2|^{e_X/2}$ times a phase which depends on the signature of $X$ \cite{FHLT}. More precisely, for any finite abelian $\Pi_2$ equipped with a non-degenerate quadratic function $q$  we can find a vector space $V$, a maximal-rank lattice $L\subset V$ and a non-degenerate even integer-valued bilinear form $B:L\otimes L\ra \ZZ$ such that $\Pi_2=L^*/\im B$, and $q$ is determined by $B$ as follows:
$$
q(x)=\frac{1}{2} B^{-1}(\tilde x,\tilde x),
$$
where $x\in L^*/\im B$ and $\tilde x$ is a lift of $x$ to $L^*=\Hom(L,\ZZ)$. For a proof see \cite{Wall}. Then  the partition sum is a Gauss sum which evaluates to \cite{FHLT, MH}:
$$
Z_X(\Pi_2,q)= |\Pi_2|^{e_X/2} e^{2\pi i \sigma(B)\sigma_X/8},
$$
where $\sigma_X$ is the signature of the intersection form of $X$ and $\sigma(B)$ is the signature of the form $B$. (It is well-known that the signature of $B$ modulo $8$ is determined by $q$). Note that this implies that for non-degenerate $q$ the partition sum is $1$ on any closed oriented 4-manifold of the form $Y\times S^1$, and therefore the space of states on any closed oriented 3-manifold $Y$ is one -dimensional. Nevertheless, the theory is not entirely trivial, because this one-dimensional vector space may be a nontrivial representation of the mapping class group of $Y$.

\section{Yang-Mills on a lattice and discrete theta-angles}

We now discuss how to construct a lattice Yang-Mills theory which flows to a particular confining TQFT in the infrared limit. The usual lattice formulation of Yang-Mills theory with gauge group $G$ involves $G$-valued variables on 1-cells of a cubic or simplicial complex $K$. This formulation does not have a room to incorporate the information about the magnetic gauge group $\Pi_2$ or discrete theta-angles. There is no satisfactory lattice definition of the ordinary continuous theta-angle either, since the instanton number is not well-defined. Below we show how to incorporate the discrete theta-angles only.

To get a more general model, one can couple lattice Yang-Mills theory to a lattice TQFT of the sort discussed above \cite{AST,GK}. In this section we show how to do it for any microscopic gauge group $G$, so that the lattice model has a well-defined 't Hooft flux taking values in a subgroup $\Pi_2$ of $\pi_1(G)$. The lattice model also has discrete theta-angles which are labeled precisely by $U(1)$-valued quadratic functions on $\Pi_2$. We conjecture that this lattice model flows to the confining TQFT with magnetic gauge group $\Pi_2$ and the same discrete theta-angles. 

The construction is very simple and not even particularly new \cite{MP,GK}. Instead of using $G$-valued variables, we use $H$-valued variables, where $t:H\ra G$ is a cover of $G$ such that $\ker\ t=\Pi_2$. Such a cover exists for any choice of $\Pi_2\subset \pi_1(G)$. For example, if $\Pi_2=\pi_1(G)$, $H$ is the universal cover. We also augment the model with $\Pi_2$-valued variables $h_\Sigma$ living on 2-cells of the complex and satisfying the constraint $\delta h=0$ (i.e. $h$ is a 2-cocycle). Let us also pick some faithful representation $R$ of $H$. Then the action is 
$$
S=\beta \sum_\Sigma \left(\Tr_R (h_\Sigma U_{\partial\Sigma})+h.c.\right)+S_{top},
$$
where $\beta$ is the inverse gauge coupling, $U_{\partial\Sigma}$ is the product of all $H$-valued variables along the boundary of $\Sigma$, and $S_{top}$ is a topological action which depends only on the variables $h_\Sigma$. 

The model has the usual gauge symmetry with gauge group $H$ as well as a discrete 1-form gauge symmetry parameterized by  $\{h_\gamma\} \in C^1(K,\Pi_2)$. Under the latter gauge symmetry the fields transform as follows:
$$
U_\gamma\mapsto h_\gamma^{-1} U_\gamma,\quad h_\Sigma\mapsto h_\Sigma \prod_{\gamma\in\partial\Sigma} h_\gamma.
$$

Wilson loops for a representation $R$ are invariant under the 1-form gauge symmetry if and only if $\Pi_2$ acts trivially on $R$. That is, if and only if $R$ is a representation of $G$. We interpret this as confinement of the subgroup $\Pi_2$ on the lattice scale, leaving behind a gauge theory with gauge group $G$. The advantage of this formulation is that since the lattice variables include a 2-cocycle $h$, one can define an observable corresponding to the 't Hooft flux, namely the cohomology class of $h$. We can also define discrete theta-angles by letting $S_{top}$ to be the most general topological action for $h$. As discussed above, such actions are classifies by quadratic functions on $\Pi_2$ with values in $U(1)$.

Discrete theta-angles and their connection with the Pontryagin square have been discussed in \cite{AST} in the context of continuum Yang-Mills theory. Now we see that the lattice formulation is able to capture all these discrete parameters via $S_{top}$, but apparently not the usual continuous theta-angle. For example, for $G=SO(3)$ the discrete theta-angle takes values in $\ZZ_4$. The continuum Yang-Mills theory also has the usual continuous theta-angle which couples to the 1st Pontryagin class of the gauge bundle and takes values in $\RR/2\pi\ZZ$. As explained in \cite{AST}, these two parameters can be combined in a single real parameter which is periodically identified\footnote{If the 4-manifold has a spin structure, then the period is $4\pi$ rather than $8\pi$ \cite{AST}.} with period $8\pi$. In the lattice formulation only the values of the parameter which are integer multiples of $2\pi$ are allowed.

In the most common case when the 4-manifold is a 4-torus, the construction of $S_{top}$ can be greatly simplified. Consider the most nontrivial case: $\Pi_2=\ZZ_r$ with $r$ even. Since the integral homology of $T^4$ is torsion-free, every cocycle in $C^2(K,\Pi_2)$ can be lifted to an integral 2-cocycle. In such a situation the Pontryagin square of $h$ is merely the square of the integral lift of $h$ modulo $2r$ (see Appendix). It is easy to see that it is independent of the lift. The topological action thus becomes
$$
S_{top}(h)=\frac{2\pi i q}{2r} (\tilde h\cup \tilde h)[T^4],
$$
where $\tilde h\in Z^2(K,\ZZ)$ is an integral lift of $h\in Z^2(K,\ZZ_r)$, and $q$ is an integer modulo $2r$ (the discrete theta-angle).
Since the intersection form of $T^4$ is even,  only the value of $q$ modulo $r$ matters in this case. Similarly, in the case $\Pi_2=\ZZ_r$ with $r$ odd we have
$$
S_{top}(h)=\frac{2\pi i q}{r} (\tilde h\cup \tilde h)[T^4],
$$
where the discrete theta-angle $q$ is now an integer modulo $r$.

While in the presence of theta-angles the weight in the lattice partition sum is not positive, this should not lead to serious problems with Monte-Carlo simulations. Indeed, the value of $S_{top}$ depends only on the cohomology class of the cocycle $h$, not on the gauge fields $U$. If instead of  fixing theta-angles one fixes the cohomology class of $h$, the topological phase factors out, and the remaining weight is real and positive (but depends on $h$). Having computed the partition function for all possible $[h]\in H^2(T^4,\Pi_2)$, one can then evaluate the partition function for all possible discrete theta-angles. For example, for $\Pi_2=\ZZ_2$ there are only 64 choices of $[h]$.

\section*{Appendix: Pontryagin square}

Let $K$ be a simplicial complex and $\Pi$ be a finite abelian group. In the simplest case $\Pi=\ZZ_r$ with $r$ even, the Pontryagin square is a cohomological operation which maps an element $f\in H^p(K,\ZZ_r)$, $p$ even, to an element $\frp f\in H^{2p}(K,\ZZ_{2r})$. It is easiest to define it if the homology group $H_{p-1}(K,\ZZ)$ is torsion-free. Then every $p$-cocycle modulo $r$ can be lifted to an integral $p$-cocycle. If $\tilde f$ is a lift of $f$, we define 
$$
\frp f= \tilde f\cup \tilde f \ {\rm mod} 2r.
$$
It is easy to see that this is well-defined (i.e. independent of the choice of the lift). 

In general one has to proceed as follows \cite{Whitehead_polyhedra}. Recall that the cup product of integral simplicial cochains is not graded-commutative. Nevertheless, the cup product in cohomology is. The way it works is as follows. Given $f\in C^p(K)$ and $g\in C^q(K)$ one has:
$$
f\cup g-(-1)^{pq} g\cup f=(-1)^{p+q-1}\left(\delta (f\cc g)-\delta f\cc g-(-1)^p f\cc \delta g\right).
$$
where $\cc$ is a new bilinear operation of degree $-1$. Let $\frp f=f\cup f+ f\cc \delta f$. If $\delta f=0$, this is simply the cup product of $f$ with itself, and therefore also an integral cocycle. But suppose $f$ is a cocycle modulo $r$ with $r$ even, i.e. $\delta f=r u$ for some $u\in C^{p+1}(K)$. Let us also assume that $p$ is even. Then
$$
\delta(\frp f)=2r f\cup u+r^2 u\cc u.
$$
Thus $\frp f$ is a degree $2p$ cocycle modulo $2r$. One can further show that $\frp f\in H^{2p}(K,\ZZ_{2r})$ is well-defined (i.e. does not change if one replaces $f\mapsto f+\delta h$ or $f\mapsto f+r g$ for some $g\in C^p(K)$). It also satisfies
$$
\frp (f_1+f_2)-\frp f_1-\frp f_2= 2 f_1\cup f_2.
$$
Note that while $f_1\cup f_2$ is defined only modulo $r$, $2f_1\cup f_2$ is defined modulo $2r$, as required. Thus $\frp$ is a quadratic refinement of the bilinear form on $H^p(K,\ZZ_{r})\times H^p(K,\ZZ_{r})$ with values in $H^{2p}(K,\ZZ_{2r})$ given by twice the cup product. Equivalently, $\frp f$ provides a canonical way to lift the class $f\cup f\in H^{2p}(K,\ZZ_{r})$ to a class in $H^{2p}(K,\ZZ_{2r})$.

It is convenient to extend the operation $\frp$ to general coefficient groups. If $\Pi$ is an abelian group, let $\Gamma(\Pi)$ be the universal quadratic group of $\Pi$. By definition, this is an abelian group equipped with a quadratic function $\gamma: \Pi\ra \Gamma(\Pi)$ such that any quadratic function $\Pi\ra A$ with values in an abelian group $A$ factors through $\gamma$. We want to define a quadratic function $\frp: H^p(K,\Pi)\ra H^{2p}(K,\Gamma(\Pi))$ which refines the bilinear form
$$
H^p(K,\Pi)\times H^p(K,\Pi)\ra H^{2p}(K,\Pi\otimes \Pi),\quad (f,g)\mapsto 2 f\cup g.
$$
Such a refinement, if it exists, is unique. For $\Pi=\ZZ_r$ with odd $r$, it is easy to see that $\Gamma(\Pi)=\ZZ_r$, so we can simply set $\frp f=f\cup f$. For $\ZZ_r$ with even $r$ we already defined $\frp$. Then we extend the definition to arbitrary finite abelian groups using the property (\ref{cross}) and by requiring that the following property holds for all $\Pi$:
$$
\frp(f_1+\ldots +f_n)=\sum_i \frp(f_i)+\sum_{i<j} 2f_i\cup f_j.
$$
Thus we obtain a functorial way to square a class in $H^p(K,\Pi)$ and get a class in $H^{2p}(K,\Gamma(\Pi))$. 

We are interested mostly in the case $p=2$. Then the Pontryagin square can be thought of as a distinguished element in $H^{4}(B^2\Pi,\Gamma(\Pi))\simeq \Hom(\Gamma(\Pi),\Gamma(\Pi))$. In fact, it corresponds to the identity element in $\Hom(\Gamma(\Pi),\Gamma(\Pi))$, because the latter also provides a quadratic refinement of twice the cup product.

An explicit formula for the product $\cc$ in the case $p=2$ and $q=3$ (the only case we need) is
$$
(f\cc g)(v_0 v_1 v_2 v_3 v_4)=f(v_0 v_3 v_4) g(v_0 v_1 v_2 v_3)+f(v_0 v_1 v_4) g(v_1 v_2 v_3 v_4).
$$
Here it is assumed that all vertices of $K$ have been ordered, and $v_0<v_1 <v_2 <v_3<v_4$ are vertices of a 4-simplex in $K$.


\begin{thebibliography}{99}

\bibitem{two} A.~Kapustin, R.~Thorngren, in preparation.

\bibitem{GK} S.~Gukov and A.~Kapustin, ``Topological Quantum Field Theory, Nonlocal Operators, and Gapped Phases of Gauge Theories,''
  arXiv:1307.4793 [hep-th].
  
\bibitem{Quinn} F.~Quinn, ``Lectures on axiomatic topological quantum field theory,'' in: {\it Geometry and Quantum Field Theory, Park City, UT, 1991,} IAS/Park City Math. Ser. {\bf 1}, 323-453, AMS, 1995. 
  
\bibitem{FHLT} D.~S.~Freed, M.~J.~Hopkins, J.~Lurie and C.~Teleman, ``Topological Quantum Field Theories from Compact Lie Groups,''
  arXiv:0905.0731 [math.AT].
  

\bibitem{AST} O.~Aharony, N.~Seiberg and Y.~Tachikawa, ``Reading between the lines of four-dimensional gauge theories,'' arXiv:1305.0318 [hep-th].

\bibitem{GNO} P.~Goddard, J.~Nuyts and D.~I.~Olive, ``Gauge Theories and Magnetic Charge,''
  Nucl.\ Phys.\ B {\bf 125}, 1 (1977).

\bibitem{KapWH} A.~Kapustin, ``Wilson-'t Hooft operators in four-dimensional gauge theories and S-duality,''
  Phys.\ Rev.\ D {\bf 74}, 025005 (2006)
  [hep-th/0501015].
  
\bibitem{DW} R.~Dijkgraaf and E.~Witten, ``Topological Gauge Theories and Group Cohomology,''
  Commun.\ Math.\ Phys.\  {\bf 129}, 393 (1990).
  
  
\bibitem{Hatcher} A.~Hatcher, ``Algebraic Topology,'' Cambridge University Press, 2001.

\bibitem{EM} S.~Eilenberg, S.~MacLane, ``On the groups $H(\Pi,n)$. II. Methods of computation,'' Ann. Math. (2) {\bf 60} (1954) 49.

\bibitem{Whitehead_certain} J.~H.~C.~Whitehead, ``A certain exact sequence,'' Ann. Math. (2) {\bf 62} (1950) 51.


\bibitem{Whitehead_polyhedra} J.~H.~C.~Whitehead, ``On simply connected, 4-dimensional polyhedra,'' Comm. Math. Helv. {\bf 22} (1949) 48.
  
\bibitem{Wall} C.~T.~C. Wall, ``Quadratic forms on finite groups and related topics,'' Topology {\bf 2} (1964) 281. 

\bibitem{MH} J.~Milnor,~D. Husemoller, ``Symmetric bilinear forms,'' Springer, 1973.

\bibitem{MP}  G.~Mack and V.~B.~Petkova, ``Z2 Monopoles In The Standard Su(2) Lattice Gauge Theory Model,'' 
  Z.\ Phys.\ C {\bf 12}, 177 (1982).
  
  


\end{thebibliography}
\end{document}